\documentclass[twocolumn,tighten]{aastex631}

\begin{document}

\shortauthors{Luhman}
\shorttitle{Candidate Substellar Members of the ONC}

\title{Candidates for Substellar Members of the Orion Nebula Cluster from 
JWST/NIRCam\footnote{Based on observations made with the Gaia mission
and the NASA/ESA/CSA James Webb Space Telescope.}}

\author{K. L. Luhman}
\affiliation{Department of Astronomy and Astrophysics,
The Pennsylvania State University, University Park, PA 16802, USA;
kll207@psu.edu}
\affiliation{Center for Exoplanets and Habitable Worlds, The
Pennsylvania State University, University Park, PA 16802, USA}

\begin{abstract}

In 2022, the James Webb Space Telescope (JWST) obtained 1--5~\micron\ images
of the center of the Orion Nebula Cluster (ONC).
I have analyzed these data in an attempt to search for substellar members 
of the cluster. Using a pair of color-color diagrams, 
I have identified $>$200 brown dwarf candidates that lack spectral 
classifications. Some of the candidates could be protostars (either
stellar or substellar) given their very red colors.
Based on the age of the ONC and the photometry predicted by theoretical 
evolutionary models, the faintest candidates could have masses of 
1--2~$M_{\rm Jup}$. This sample of candidates may prove to be valuable 
for studying various aspects of young brown dwarfs, including their mass
function and minimum mass.  However, spectroscopy is needed to confirm the 
membership (via signatures of youth) and late spectral types of the candidates.
Finally, I note that most of the ``Jupiter-mass binary objects" that
have been previously identified with these JWST images
are absent from my sample of candidates because their colors are indicative of 
reddened background sources rather than young brown dwarfs or their
photometry is inadequate for assessing their nature because of very low 
signal-to-noise ratios and/or detections in only a few bands.

\end{abstract}

\section{Introduction}
\label{sec:intro}

The Orion Nebula Cluster \citep[ONC,][]{mue08,ode08} is one of the best
available sites for measuring a statistically significant mass function
of brown dwarfs down to their minimum mass because it is young 
\citep[$\tau\sim1$--5 Myr,][]{jef11}, nearby \citep[$d=390\pm2$ pc,][]{mai22},
rich ($N\sim$2000 members), and compact ($D\sim0\fdg5$).
Brown dwarf candidates have been identified in the ONC using
optical and near-infrared cameras on ground-based telescopes
\citep{mcc94,sim99,luc00,hil00,mue02,luc05,rob10,dar12,dra16,mei16}
and the Hubble Space Telescope \citep{luh00,and11,rob20,gen20}, reaching
magnitudes that should correspond to a few Jupiter masses.
Ultimately, spectroscopy is required for confirming the youth and late spectral 
types of brown dwarf candidates in star-forming regions, which has been
acquired for a small fraction of those found in the ONC
\citep{hil97,luc01,luc06,sle04,rid07,wei09,hil13,ing14,luh24o}.

As the most sensitive infrared (IR) telescope to date, the James Webb Space 
Telescope \citep[JWST,][]{gar23} is capable of extending surveys for
brown dwarfs in the ONC down to the mass of Jupiter.
The small field of view of JWST is well matched to the compact spatial
distribution of ONC members while its high angular resolution is vital
for the detection of faint sources within the bright emission
from the Orion Nebula. A few months after JWST began science operations,
\citet{mcc23} used it to obtain 1--5~\micron\ images in the center of the ONC. 
They reported the detection of several hundred candidates for planetary mass 
brown dwarfs, some of which resided in close pairs that were interpreted as 
possible binary systems and that were named ``Jupiter-mass binary objects"
\citep[JuMBOs,][]{pea23}.
In this paper, I present an independent analysis of those images in an attempt 
to identify a reliable sample of brown dwarf candidates and to assess the 
validity of the previous candidates.

\section{Reduction of NIRCam Data}

The ONC images from \citet{mcc23} were taken with NIRCam on JWST 
\citep{rie05,rie23}. The camera simultaneously observes 
nearly identical areas of sky through short and long wavelength channels 
(0.7--2.4 and 2.4--5~\micron). The short/long wavelength channels utilize
eight/two 2040$\times$2040 detector arrays with pixel sizes of
$0\farcs031/0\farcs063$. The detectors cover two
$2\farcm2\times2\farcm2$ fields that are separated by $43\arcsec$.
The set of detectors in each of the two fields is referred to as a module.
For each of the two modules, $5\arcsec$ gaps separate the four detectors 
in the short wavelength channel.

The NIRCam observations were performed through guaranteed time observation 
(GTO) program 1256 (PI: M. McCaughrean) in September and October of 2022,
as described by \citet{mcc23}. I summarize the aspects of 
the observations that are most relevant to my reduction of the data.
The images were obtained in 12 filters between 1--5~\micron, consisting of
F115W, F140M, F162M, F182M, F187N (Pa-$\alpha$), and F212N (H$_2$ rovibrational
transition) in the short wavelength channel and F277W, F300M, F335M, F360M, 
F444W, and F470N (H$_2$ rotational transition) in the long wavelength channel.
The images in F115W and F444W were taken simultaneously in a mosaic
pattern that contained 14 pointings and covered $11\farcm7\times7\farcm5$.
The data in the remaining filters were obtained with a mosaic pattern that
contained 10 pointings and covered $11\arcmin\times7\farcm5$.
For a given pointing in a mosaic, the dithered exposures for all
selected filters were collected within a single visit, resulting in
a total of 24 visits. In the mosaic pattern for F115W/F444W, there
was significant overlap between the images of neighboring pointings/visits.
For the other filters, most of the overlap was confined to within visits 1--4,
5--8, and 9/10, and there was little overlap between those three sections
of the mosaic.

To reduce the NIRCam data, I began by retrieving the {\tt uncal} files 
from the Mikulksi Archive for Space Telescopes (MAST):
\dataset[doi:10.17909/vjys-x251]{http://dx.doi.org/10.17909/vjys-x251}.
The data were processed with version 
\dataset[1.15.1]{https://doi.org/10.5281/zenodo.7829329}
of the JWST Science Calibration pipeline under context jwst\_1242.pmap.
The {\tt calwebb\_detector1} pipeline module was used to apply
detector-level corrections to the {\tt uncal} files. The resulting count
rate images ({\tt rate}) were processed with the {\tt calwebb\_image2} module,
producing calibrated unrectified exposures ({\tt cal}).

The {\tt tweakreg} routine within the {\tt calwebb\_image3} module is 
responsible for image registration. For 10 of the 12 filters, the images 
have insufficient overlap for internal registration across the entire mosaic. 
Therefore, I used high precision astrometry from the third data 
release of the Gaia mission \citep{gaia16b,bro21,val23} as an 
external reference catalog in the first stage of the registration process.
The F470N images contain the largest number of nonsaturated detections of 
Gaia sources, so I used {\tt tweakreg} 
to perform (1) relative alignment
of the dithered F470N exposures for a given module (a single detector) 
and visit based on their detected sources and (2) an absolute
alignment of that set of dithered exposures based on nonsaturated detections
of sources from Gaia DR3. The aligned F470N images from all visits were
processed simultaneously by {\tt calwebb\_image3} with {\tt tweakreg} 
deactivated to produce a mosaic. 
I applied {\tt tweakreg} to the dithered exposures in F115W for a given
module (four detectors) and visit using sources from the F470N mosaic as the
absolute reference catalog. This approach does not solve for separate
offsets for the individual detectors, and instead assumes that the pipeline's 
relative world coordinate systems (WCSs) for the four detectors in a
module are accurate. The aligned images in F115W were combined into 
a mosaic. For each of the remaining 10 filters, I aligned the dithered
exposures for a given module and visit using sources from the F115W mosaic
as the absolute reference catalog. 

When performing {\tt calwebb\_image3} with the option for background matching, 
some areas of the resulting mosaics with brighter nebular emission 
exhibited discontinuities in the background level that reflected improper 
matching. These discontinuities were largely absent when that routine 
was deactivated, so that was done for the final mosaics.

After constructing the mosaics
for all filters and generating initial catalogs of sources, I calculated 
the offsets in celestial coordinates between pairs of catalogs.
For most filter pairs, the offsets clustered tightly with residuals of a few
milliarcseconds. However, offsets involving F162M as one of the filters
spanned a much larger range ($\sim$20 mas).
I reprocessed the F162M data with {\tt tweakreg} applied to the dithered
exposures of individual detectors. The resulting mosaic did not exhibit
the previous discrepancy in the offsets relative to other filters, 
which suggests the presence of a significant error in the pipeline's
relative WCSs of the detectors when processing images in F162M.

The reduced mosaic images for the 12 filters are available at
\dataset[doi:10.5281/zenodo.13824020]{https://doi.org/10.5281/zenodo.13824020}.

\begin{figure}
\epsscale{1.2}
\plotone{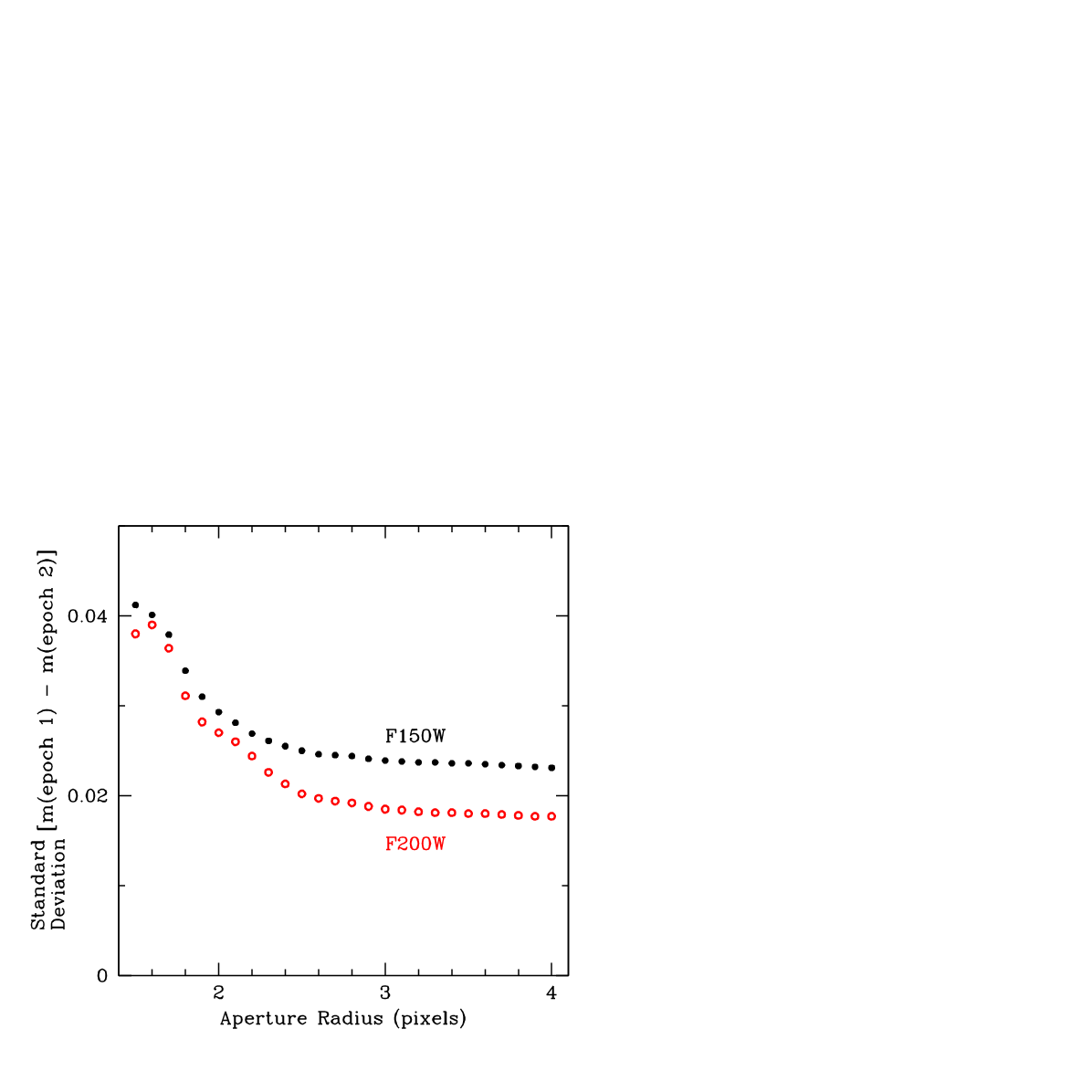}
\caption{Standard deviation of the differences in NIRCam photometry between
two epochs of images versus the aperture radius for the photometry.
These data were measured for bright stars in archival images of a field in 
the LMC obtained in the F150W and F200W filters.}
\label{fig:sig}
\end{figure}

I identified sources in the mosaic images using a combination of the 
{\tt starfind} routine in IRAF and visual inspection of the images in
the manner described by \citet{luh24i} for NIRCam images of IC 348.
In that study, the default photometry utilized an aperture radius of 3 pixels,
which was selected to be somewhat larger than the FWHM of the point spread
function ($\sim2$ pixels at 2/4~\micron\ in the
short/long wavelength channels). I have 
refined the selection of an aperture radius for the ONC images by 
identifying the smallest radius that minimizes the residuals
between measurements at different epochs for a given sample of stars.
For this exercise, I selected NIRCam calibration images of a field in the
Large Magellanic Cloud (LMC) that were obtained in May and June of 2022 
(program 1074, PI: M. Robberto):
\dataset[doi:10.17909/epes-nx15]{http://dx.doi.org/10.17909/epes-nx15}.
These data were taken in F150W and F200W and contain a large number of point 
sources that are not saturated and have high signal-to-noise ratios (S/Ns).
Using the routine {\tt phot} in IRAF, I measured photometry for
sources in the LMC images for radii that ranged from 1.5 to 4 pixels
at increments of 0.1 pixel. The inner and outer radii of the sky annuli 
were 4 and 8 pixels, respectively. For the brightest $\sim1$ mag of 
nonsaturated stars, I calculated the standard deviation of the differences
in photometry between the two epochs for each aperture radius.
The resulting standard deviations are plotted versus radius in 
Figure~\ref{fig:sig}. For both filters, the standard deviations
reach a minimum at radii of $\gtrsim$2.5 pixels. 
I also performed the same analysis on F150W and F200W images of the LMC
field from November 2022 and May 2023 (program 1476, PI: M. Boyer),
which produced similar results. Therefore, I have selected an 
aperture radius of 2.5 pixels for the default photometry in the ONC images.
The sky annulus is the same as the one applied to the LMC data.

As done in \citet{luh24i}, I have used the differences in photometry between
aperture radii of 2 and 4 pixels to identify candidates for resolved galaxies
in the ONC data. I have visually inspected the NIRCam images of those sources,
rejecting those that have the morphologies of galaxies and retaining the 
remaining sources (e.g., close pairs of stars, reflection nebulae).

I have not utilized the photometry in 
F212N, F300M, F335M, and F470N since these bands are not useful for identifying
brown dwarfs or are redundant with other filters that are more sensitive.
The band-merged catalog for the remaining eight filters contains $\sim$1500
sources that are not saturated in at least one medium-/broad-band filter.
Saturation occurs near magnitudes of $\sim$15.5--16 in F115W, F140M, 
F162M, F182M, and F277W, $\sim$12 in F187N, and $\sim$14 in F360M and F444W.
Since most sources in the NIRCam images of the ONC are at least moderately red
because of reddening from the Orion molecular cloud,
sources that are slightly fainter than the saturation 
limit in a given band are often saturated at longer wavelengths.
Measurements with errors greater than 0.2 mag are omitted
from my analysis and the source catalogs. Thus, a source that lacks
photometry in a given band may have a detection, but with a low S/N.

\section{Analysis of NIRCam Photometry}

\subsection{Point Sources with Spectral Classifications}

ONC sources that have been previously found to have late spectral types
can help to guide the selection of substellar candidates in the NIRCam data.
I have compiled all available spectral types for NIRCam point sources that have
photometry in at least one medium-/broad-band filter.
The resulting catalog contains 79 sources and is presented in 
Table~\ref{tab:m}. Extended sources that have classifications are discussed 
in Section~\ref{sec:e}.
One of the point sources has been classified as a field dwarf, so it is 
omitted from Table~\ref{tab:m}. Some of the remaining stars in that tabulation 
lack clear indicators of age in their spectra, and may not be members of 
the ONC \citep{sle04}. In addition to spectral types, Table~\ref{tab:m} includes
source names from this work, names from a selection of previous studies 
\citep{ode94,hil97,her02,luc05,rod09,luh24o}, and celestial coordinates and 
photometry measured from the eight bands of NIRCam images that I have utilized.

\begin{figure}
\epsscale{1.2}
\plotone{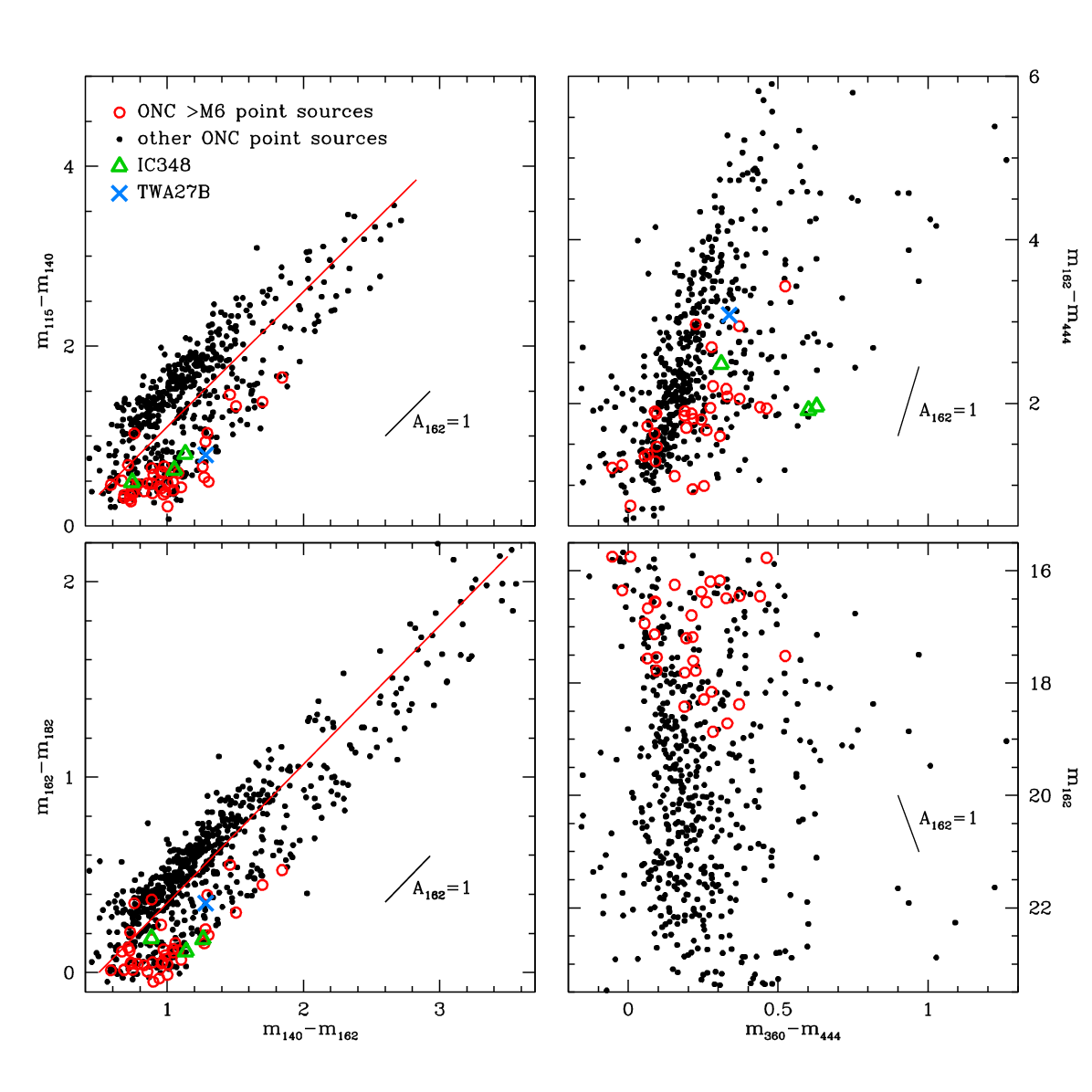}
\caption{
Color-color and color-magnitude diagrams for point sources in JWST/NIRCam 
images of the ONC (points and circles). 
Three of the diagrams include data for TWA 27B and three late-type members
of IC 348 \citep[cross and triangles,][]{luh23,luh24i}. 
The red lines in the left diagrams are reddening vectors used for selecting 
brown dwarf candidates.}
\label{fig:cmd1}
\end{figure}

Because of their broad molecular absorption bands, the spectral energy 
distributions of brown dwarfs are complex and distinctive from those
of other astronomical sources. As a result, brown dwarf candidates can be 
identified using colors if an appropriate combination of filters is available, 
even in the presence of a wide range of extinctions.  In Figure~\ref{fig:cmd1}, 
I present the two color-color diagrams for the NIRCam data in which brown 
dwarfs are most easily distinguished from stars and galaxies,
consisting of $m_{115}-m_{140}$ and $m_{162}-m_{182}$ versus $m_{140}-m_{162}$. 
At the age of the ONC, the hydrogen burning mass limit should correspond
to a spectral type of $\sim$M6.5 based on temperature scales for young stars
\citep{luh03} and theoretical evolutionary models \citep{bar98,bar15,cha23}. 
Therefore, NIRCam point sources that have classifications later than M6 and 
evidence of youth are marked with a separate symbol.  
To further illustrate the expected colors for substellar members of the ONC,
I have included in Figure~\ref{fig:cmd1} the three faintest
known members of the young cluster IC 348 \citep{luh24i} and 
the young planetary mass companion TWA 27B \citep{cha04,luh23}.  
Sources with extended emission are excluded from Figure~\ref{fig:cmd1} 
(see Section~\ref{sec:e}).

In each of the two color-color diagrams in Figure~\ref{fig:cmd1}, the
overall distribution of sources exhibits a linear trend that corresponds
to the influence of reddening from the Orion molecular cloud. 
A reddening vector for $A_{162}=1$ is indicated \citep{sch16av}.
At a given value of $m_{140}-m_{162}$, the known $>$M6 sources have
blue colors in $m_{115}-m_{140}$ and $m_{162}-m_{182}$ relative to
the richest concentration of sources in each diagram, which likely consists
of background stars and unresolved galaxies. Most of the known $>$M6 sources 
have colors that suggest low-to-moderate extinctions ($A_{162}<1$).
A few of the late-type objects have redder colors, which can reflect
higher extinctions or later spectral types.  For instance, the L dwarf TWA 27B 
likely has negligible extinction and is intrinsically very red.
In each of the color-color diagrams, I have marked a reddening vector that
captures most of the known $>$M6 objects in the ONC as well as TWA 27B and
the brown dwarfs in IC 348. These vectors will be used for selecting
brown dwarf candidates in Section~\ref{sec:c}.
For the few $>$M6 sources that appear above those vectors, the previous spectra 
have low S/Ns, so their classifications are uncertain.
I note that in $m_{162}-m_{277}$ versus $m_{140}-m_{162}$, brown dwarfs
have a roughly similar pattern as in the aforementioned color-color diagrams,
except that some L dwarfs like TWA 27B overlap with the locus of reddened
stars and galaxies, making it less useful for the selection of brown dwarf
candidates.

I have included an additional color-color diagram in Figure~\ref{fig:cmd1}, 
consisting of $m_{162}-m_{444}$ versus $m_{360}-m_{444}$. A diagram of this 
kind was utilized in a NIRCam survey for brown dwarfs in IC 348 \citep{luh24i}.
Most of the known $>$M6 sources are only moderately red in $m_{162}-m_{444}$ 
and $m_{360}-m_{444}$, but ONC brown dwarfs can appear redder if they have later
spectral types, higher reddenings, or IR excess emission from disks. 
The faintest members of IC 348 reside in an area of the color-color diagram
that has little contamination from background sources.

Brown dwarf surveys in star-forming regions often use color-magnitude diagrams
to select candidates that have the expected magnitudes for a given color, i.e.,
appearing near the sequence observed or predicted for members. 
As done in IC 348, I have employed a diagram of $m_{162}$ versus 
$m_{360}-m_{444}$ for the ONC data, which is shown
in Figure~\ref{fig:cmd1}. Once again, redder colors in this diagram can
reflect later spectral types, higher reddenings, or excess emission from disks.
The known $>$M6 sources in the ONC span a small range of colors. 
The intrinsic values of this color are similar for late M and L dwarfs, and grow
significantly redder only with the onset of absorption from CH$_4$ or other 
hydrocarbons within the F360M band, as in the case for two faintest objects 
in IC 348 \citep{luh24i}.

\subsection{Point Source Brown Dwarf Candidates}
\label{sec:c}

For my sample of brown dwarf candidates in the ONC, I initially selected point 
sources that (1) appear below either of the red boundaries in the diagrams
of $m_{115}-m_{140}$ and $m_{162}-m_{182}$ versus $m_{140}-m_{162}$
from Figure~\ref{fig:cmd1}, (2) do not appear above the boundary
in either of those diagrams, and (3) lack spectral classifications.
The resulting sample contains 245 sources. 

The Chandra Orion Ultradeep Project (COUP) performed deep X-ray observations
of a field in the ONC that fully encompasses the NIRCam mosaics
\citep{get05a,get05b}. Forty-three of the brown dwarf candidates identified
with NIRCam have COUP counterparts within roughly twice the positional errors
from COUP.  Three of those sources were classified as extragalactic based on
the X-ray data \citep{get05b}, so they are excluded from my sample of
candidates.  Most of the remaining X-ray-detected candidates are very red
in NIRCam and are saturated in the bands at longer wavelengths. These sources
may be protostars. In general, most of the reddest brown dwarf candidates
($m_{140}-m_{162}>2.5$) are among the brighter candidates ($m_{162}<19$) 
and may be protostars (either low-mass stars or brown dwarfs).

\begin{figure}
\epsscale{1.2}
\plotone{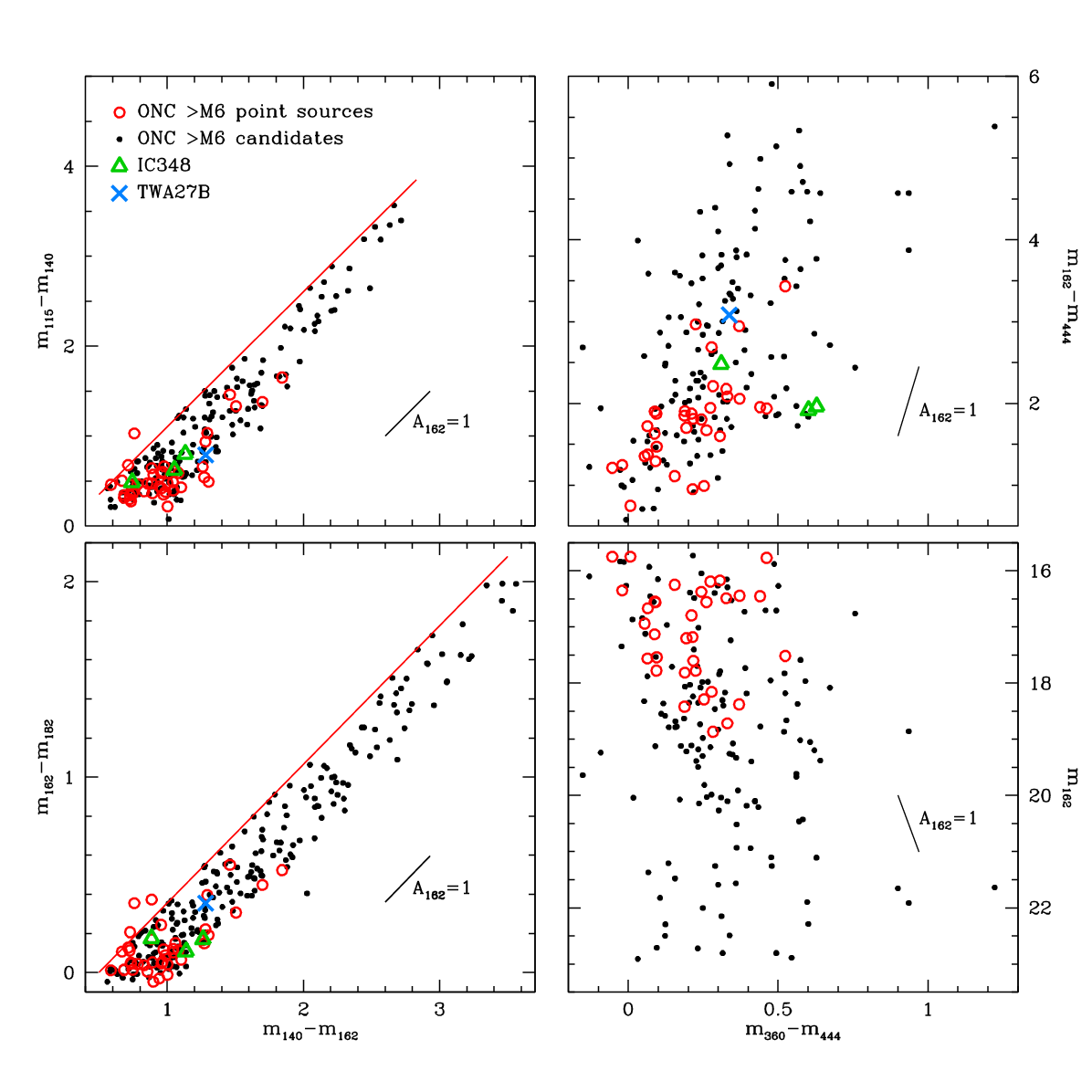}
\caption{
Color-color and color-magnitude diagrams for point source brown dwarf
candidates that are selected based on their positions below the red
boundaries in the left diagrams (Table~\ref{tab:c}).}
\label{fig:cmd2}
\end{figure}

The 242 brown dwarf candidates are presented in Table~\ref{tab:c}
and are plotted in the color-color
and color-magnitude diagrams in Figure~\ref{fig:cmd2}. The samples in
the four diagrams are not identical because of nondetections and saturation.
For instance, 101 of the candidates are absent from the color-magnitude diagram 
because of either saturation or low S/N in F360M and F444W.
Strong Pa-$\alpha$ emission from intense accretion or an ionization front
could contaminate the F182M photometry and place a brown dwarf above the 
boundary in $m_{162}-m_{182}$ versus $m_{140}-m_{162}$, but this scenario
appears to be rare among point-like brown dwarfs based on the sample of
sources with spectral classifications.

In the third color-color diagram in Figure~\ref{fig:cmd2}, $m_{162}-m_{444}$ 
versus $m_{360}-m_{444}$, some of the brown dwarf candidates overlap with 
the known $>$M6 objects. Several candidates also appear near the faintest
members of IC 348, which suggests that they could have especially late
spectral types. Other candidates are much redder in $m_{162}-m_{444}$
than the known late-type sources, which could be unresolved galaxies
or brown dwarfs that have high extinctions or IR excess emission.

In the diagram of $m_{162}$ versus $m_{360}-m_{444}$, most of the candidates 
have positions that are roughly consistent with the sequence of known late-type 
objects or its extension to fainter magnitudes. A minority of candidates seem 
less likely to be ONC brown dwarfs based on their faint magnitudes and
blue colors. However, in the F360M and F444W images, some of those objects 
are located within diffraction spikes from bright stars or bright background 
emission, so their values of $m_{360}-m_{444}$ may not be reliable.
For that reason, and because the shape of the sequence of young brown dwarfs
at the faintest magnitudes in NIRCam has not been measured in any region, 
the faint blue sources are retained in my sample of candidates.

My selection criteria for brown dwarf candidates require photometry in
F115W, F140M, and F162M, or in F140M, F162M, and F182M. 
Some of the known late-type sources in the ONC are saturated in these bands.
Brown dwarf candidates at those brighter magnitudes have been identified 
with previous imaging \citep[e.g.,][]{rob20}.
Meanwhile, some NIRCam sources are red enough that they have nondetections
in too many bands at shorter wavelengths for the application of my criteria.
For those reddest sources, the available bands of
photometry are insufficient for distinguishing substellar members of the ONC 
from other types of objects, such as protostars and galaxies.

For the background levels found in most areas of the sky, unreddened
brown dwarfs are detected with roughly similar S/N's in the short and long
wavelength channels of NIRCam (e.g., F162M and F444W) at a given exposure 
time\footnote{https://jwst.etc.stsci.edu/}. However, the center of the
ONC has unusually bright background emission, and that emission is brighter
at longer wavelengths. As a result, the bands in the short wavelength channel 
that I have used for the selection of brown dwarf candidates offer better 
sensitivity for low-to-moderate extinctions, as well as better discrimination 
between brown dwarfs and contaminants. It is likely that the only ONC brown 
dwarfs that are detected in the long wavelength channel alone are those with 
high extinctions.

\subsection{Proplyds and Other Extended Sources}
\label{sec:e}

A member of a star-forming region can have extended emission in the form
of scattered light from an edge-on disk or the ambient molecular cloud.
In a dense cluster like the ONC, close proximity to an O star can also 
result in the formation of an ionization front surrounding a star's disk,
which is known as a ``proplyd" \citep{ode93,ode94,ode96}.
These types of extended emission are relevant to my search for brown dwarfs
in the ONC because they serve as evidence of membership in the cluster.
In addition, young stars that are dominated by scattered light can have
unusual colors, so brown dwarfs with extended emission may not satisfy
the photometric selection criteria from Section~\ref{sec:c}. 
Therefore, I have performed a separate analysis of extended sources.

To search for brown dwarf candidates that have extended emission, I have
compiled all sources in the NIRCam field that are known to have extended 
emission that is indicative of a young star based on Hubble images
\citep{ode93,ode94,ode96,mcc96,bal00,ric08} and have searched for new objects
of this kind in the NIRCam images. In Table~\ref{tab:e}, I present
a catalog of 64 extended sources that have photometry in at least one 
medium-/broad-band filter from NIRCam. The previously known proplyds in 
Table~\ref{tab:e} have source names from \citet{ode94} and \citet{ric08}.
I have omitted a few objects that are diffuse in all bands and that lack 
a centrally concentrated source of emission that is suitable for aperture
photometry. One of those omitted objects is [LRT2005] 057$-$305, which has
a new edge-on disk.  
Some of the objects in Table~\ref{tab:e} have spectral classifications,
some of which are uncertain. The catalog includes two proplyds in which the
central sources have been recently classified as M6.5 and M7.5 through JWST 
spectroscopy, placing them near and below the hydrogen burning limit, 
respectively \citep{luh24o}.
Similar observations for all of the remaining sources in Table~\ref{tab:e}
would be useful for determining whether they are likely to be brown dwarfs.

\begin{figure}
\epsscale{1.2}
\plotone{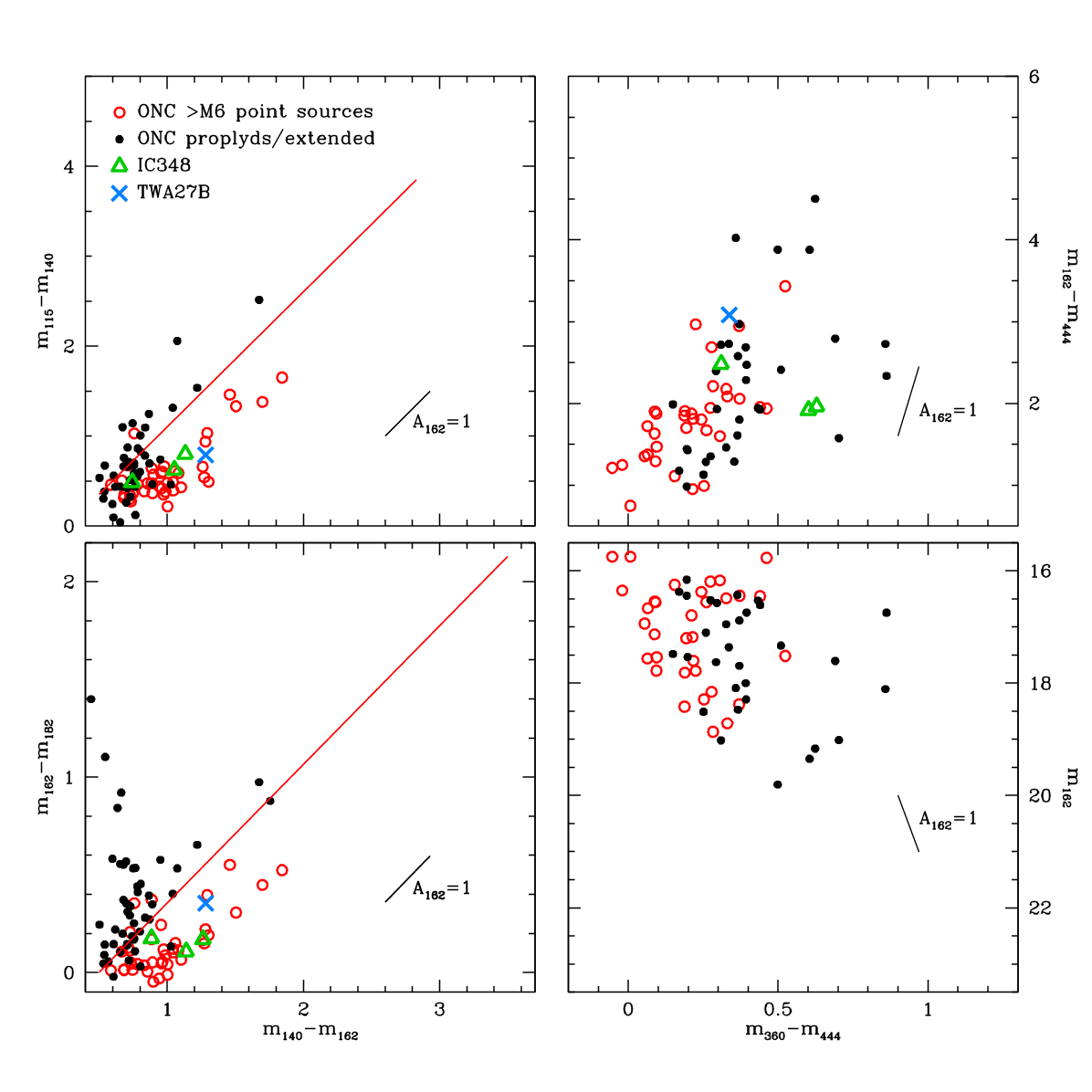}
\caption{
Color-color and color-magnitude diagrams for proplyds and other sources with
extended emission (Table~\ref{tab:e}).}
\label{fig:cmd3}
\end{figure}

The aperture photometry for the extended sources should be treated with
caution, but it is still useful to examine their colors. 
In Figure~\ref{fig:cmd3}, they are plotted in the same color-color
and color-magnitude diagrams that were utilized in the previous two sections.
Some objects are absent from a given diagram because of
saturation or nondetection.
In the diagram of $m_{115}-m_{140}$ versus $m_{140}-m_{162}$,
roughly half of the extended sources appear below the threshold that was
used for selecting brown dwarf candidates. Only a few sources are below
the threshold in $m_{162}-m_{182}$ versus $m_{140}-m_{162}$, but that
can be explained by contamination of the F182M photometry by Pa-$\alpha$
emission, as mentioned earlier. The positions of the extended sources
in the remaining two diagrams in Figure~\ref{fig:cmd3} are also consistent
with substellar members of the ONC.

\subsection{Comparison to Previous Work}

Based on their analysis of the NIRCam images, \citet{mcc23} and \citet{pea23}
reported the identification of 540 candidates for ONC members with mass 
estimates of 0.6--13~$M_{\rm Jup}$.
A subset of these objects appeared in pairs and trios with separations of
0.07--$1\arcsec$, which were interpreted as multiple systems and were named
JuMBOs. A large majority of the JuMBOs have separations that correspond 
to $>$100 AU at the distance of the ONC. Binary brown dwarfs at such wide 
separations are rare in other star-forming regions \citep{tod14}.

\begin{figure}
\epsscale{1.2}
\plotone{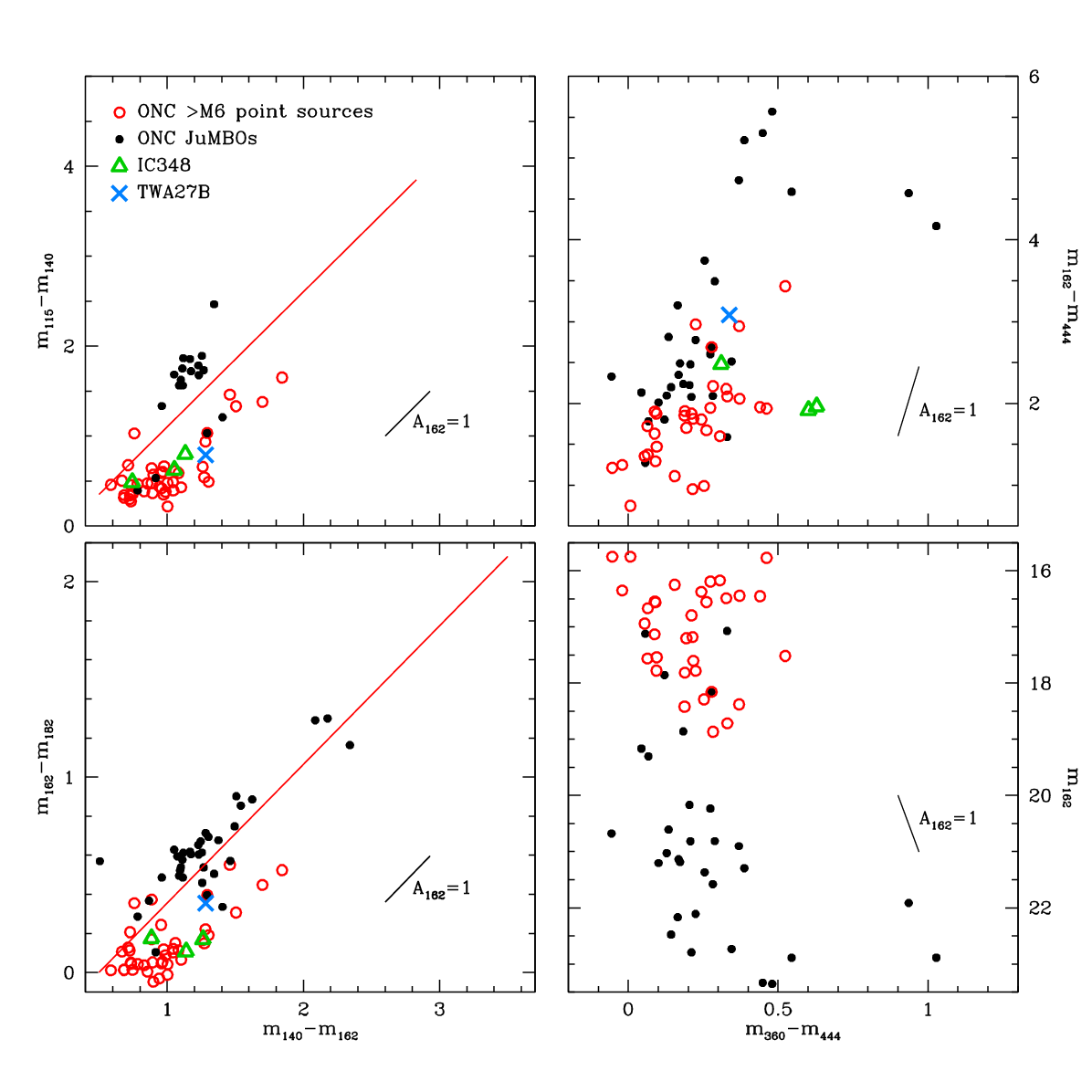}
\caption{
Color-color and color-magnitude diagrams for the JuMBO candidates
from \citet{pea23}.}
\label{fig:cmd4}
\end{figure}

Among the 86 JuMBO components from \citet{pea23}, 60 have entries in my
NIRCam catalog. For the remaining 26 components, the S/N's are too low in
all bands for useful photometry. 
The 60 components in my catalog are plotted in color-color and color-magnitude
diagrams in Figure~\ref{fig:cmd4}. Some sources lack the S/N's needed
to appear in a given diagram.
In the two color-color diagrams that I have used for selecting brown dwarf 
candidates, most of the JuMBOs have colors that are inconsistent with young
brown dwarfs and instead are indicative of background sources.
Most of them are faint and blue in the color-magnitude diagram
in Figure~\ref{fig:cmd4}, which is also suggestive of contaminants.
A few of the JuMBO components are viable brown dwarf candidates.
Source 29 from \citet{pea23} ([LAB20024] 139) has been spectroscopically
classified as young and M8 \citep{luh24o}. Its candidate companion has
a separation of $0\farcs33$, is fainter, and is in my sample of brown dwarf 
candidates. For sources 12, 23, 27, and 31 from \citet{pea23}, one component
of each pair satisfies my selection criteria. Thus, one pair from
that study is a good candidate for a substellar binary system.
\citet{mcc23} and \citet{pea23} did not present a tabulation of their
candidates for single brown dwarfs.

\citet{rod24} presented radio observations of one of the JuMBOs, source 24 
from \citet{pea23} ([HC2000] 728). The components of this pair
have a separation of $0\farcs1$ and have similar magnitudes in NIRCam images.
Unresolved spectroscopy for the pair has produced a spectral type of M5.5
\citep{sle04}, which is near the peak of the initial mass function
in star-forming regions and corresponds to a mass of 
$\sim0.1$--0.15~$M_\odot$ according to evolutionary models \citep{cha23}.
Like a number of faint ONC candidates that have been observed with 
spectroscopy \citep{sle04}, this pair is unusually faint for
its spectral type relative to members of the cluster.
The typical explanation for sources of this kind is that they are
background stars or cluster members that are observed primarily in
scattered light. Gravity-sensitive lines in the spectrum of the system
support youth \citep{sle04}, but the NIRCam images do not show extended
emission or excess emission in F444W from circumstellar dust.
In either case, the spectroscopy indicates that the pair consists of
low-mass stars.

\subsection{Mass Estimates of Brown Dwarf Candidates}

The faintest brown dwarf candidates from this work have $m_{162}\sim23$. 
Some of them appear to have little extinction since their observed colors 
are roughly similar to the intrinsic colors of young brown dwarfs. 
If they have no extinction and are located at the distance of the ONC, 
their absolute magnitudes would be $\sim15$ in F162M.
Assuming an age of 1--5 Myr for the ONC \citep{jef11}, those
faintest candidates should have masses of 1--2~$M_{\rm Jup}$ if they
are ONC members based on the photometry predicted by theoretical evolutionary
models \citep{cha23}.

\section{Conclusions}

I have analyzed JWST/NIRCam images from 1--5~\micron\ for a 
$11\arcmin\times7\farcm5$ field in the center of the ONC \citep{mcc23} 
in an attempt to identity candidates for substellar members of the cluster. 
The results are summarized as follows:

\begin{enumerate}

\item
Following the reduction of the NIRCam images and the construction of source
catalogs, I have compiled all available spectral types for point sources 
that have photometry in at least one medium-/broad-band filter of NIRCam
(Table~\ref{tab:m}).  I find that ONC sources 
with types that correspond to substellar masses ($>$M6) are distinctive from 
most background sources in a pair of color-color diagrams. 
I have used these diagrams to identify 242 brown dwarf candidates from among 
NIRCam point sources that lack spectral classifications (Table~\ref{tab:c}).
Some of these candidates could be protostars (either stellar or substellar)
based on their very red colors.

\item
Since brown dwarfs with extended emission (e.g., proplyds, edge-on disks)
may have unusual colors that would not satisfy my selection criteria,
I have compiled nonsaturated sources in the NIRCam images that are known to 
have extended emission based on Hubble images \citep[e.g.,][]{ode94,ric08} 
and I have searched for new objects of this kind in the NIRCam data, 
arriving at a sample of 64 sources (Table~\ref{tab:e}).
Although their photometry may not be reliable, 
some of the objects do have colors that are suggestive of brown dwarfs.

\item
For both the point-like and extended candidates, spectroscopy is needed to
confirm their membership (via signatures of youth) and late spectral types.
Based on the age of the ONC and the photometry predicted by evolutionary 
models \citep{cha23}, the faintest candidates could have masses of
1--2~$M_{\rm Jup}$. 

\item
Previous studies of the NIRCam data reported the detection of several 
hundred brown dwarf candidates with mass estimates below 13~~$M_{\rm Jup}$,
some of which appeared in pairs and trios that were interpreted as 
multiple systems and were named JuMBOs. 
I find that most of the JuMBO components have colors that are indicative of  
reddened background sources rather than brown dwarfs or have inadequate
photometry for assessing their nature because of very low S/Ns 
and/or detections in only a few bands.

\end{enumerate}

\begin{acknowledgments}

The JWST data were obtained from 
MAST at the Space Telescope Science Institute, which is operated by the 
Association of Universities for Research in Astronomy, Inc., under NASA 
contract NAS 5-03127. The JWST observations are associated with program 1256.
This work used data from the European Space Agency
mission Gaia (\url{https://www.cosmos.esa.int/gaia}), processed by
the Gaia Data Processing and Analysis Consortium (DPAC,
\url{https://www.cosmos.esa.int/web/gaia/dpac/consortium}).
The Center for Exoplanets and Habitable Worlds is supported by the
Pennsylvania State University, the Eberly College of Science, and the
Pennsylvania Space Grant Consortium.

\end{acknowledgments}

\clearpage

\begin{deluxetable}{ll} 
\tabletypesize{\scriptsize}
\tablewidth{0pt}
\tablecaption{Point Sources with Spectral Classifications\label{tab:m}}
\tablehead{
\colhead{Column Label} &
\colhead{Description}} 
\startdata
ID & Source name from this work \\
Name & Other source name \\
RAdeg & Right ascension (ICRS)\\
DEdeg & Declination (ICRS)\\
SpType & Spectral type \\
r\_SpType & Spectral type reference\tablenotemark{a} \\
m115mag & F115W NIRCam magnitude \\
e\_m115mag & Error in m115mag \\
m140mag & F140M NIRCam magnitude \\
e\_m140mag & Error in m140mag \\
m162mag & F162M NIRCam magnitude \\
e\_m162mag & Error in m162mag \\
m182mag & F182M NIRCam magnitude \\
e\_m182mag & Error in m182mag \\
m187mag & F187M NIRCam magnitude \\
e\_m187mag & Error in m187mag \\
m277mag & F277W NIRCam magnitude \\
e\_m277mag & Error in m277mag \\
m360mag & F360M NIRCam magnitude \\
e\_m360mag & Error in m360mag \\
m444mag & F444W NIRCam magnitude \\
e\_m444mag & Error in m444mag
\enddata
\tablenotetext{a}{
(1) \citet{wei09};
(2) \citet{ing14};
(3) \citet{luc06};
(4) \citet{rid07};
(5) \citet{sle04};
(6) \citet{luh24o};
(7) \citet{hil97}.}
\tablecomments{
The table is available in its entirety in machine-readable form.}
\end{deluxetable}

\begin{deluxetable}{ll} 
\tabletypesize{\scriptsize}
\tablewidth{0pt}
\tablecaption{Point Source Brown Dwarf Candidates\label{tab:c}}
\tablehead{
\colhead{Column Label} &
\colhead{Description}} 
\startdata
ID & Source name from this work \\
Name & Other source name \\
RAdeg & Right ascension (ICRS)\\
DEdeg & Declination (ICRS)\\
m115mag & F115W NIRCam magnitude \\
e\_m115mag & Error in m115mag \\
m140mag & F140M NIRCam magnitude \\
e\_m140mag & Error in m140mag \\
m162mag & F162M NIRCam magnitude \\
e\_m162mag & Error in m162mag \\
m182mag & F182M NIRCam magnitude \\
e\_m182mag & Error in m182mag \\
m187mag & F187M NIRCam magnitude \\
e\_m187mag & Error in m187mag \\
m277mag & F277W NIRCam magnitude \\
e\_m277mag & Error in m277mag \\
m360mag & F360M NIRCam magnitude \\
e\_m360mag & Error in m360mag \\
m444mag & F444W NIRCam magnitude \\
e\_m444mag & Error in m444mag
\enddata
\tablecomments{
The table is available in its entirety in machine-readable form.}
\end{deluxetable}

\begin{deluxetable}{ll} 
\tabletypesize{\scriptsize}
\tablewidth{0pt}
\tablecaption{Proplyds and Other Extended Sources\label{tab:e}}
\tablehead{
\colhead{Column Label} &
\colhead{Description}} 
\startdata
ID & Source name from this work \\
Name & Other source name \\
RAdeg & Right ascension (ICRS)\\
DEdeg & Declination (ICRS)\\
SpType & Spectral type \\
r\_SpType & Spectral type reference\tablenotemark{a} \\
m115mag & F115W NIRCam magnitude \\
e\_m115mag & Error in m115mag \\
m140mag & F140M NIRCam magnitude \\
e\_m140mag & Error in m140mag \\
m162mag & F162M NIRCam magnitude \\
e\_m162mag & Error in m162mag \\
m182mag & F182M NIRCam magnitude \\
e\_m182mag & Error in m182mag \\
m187mag & F187M NIRCam magnitude \\
e\_m187mag & Error in m187mag \\
m277mag & F277W NIRCam magnitude \\
e\_m277mag & Error in m277mag \\
m360mag & F360M NIRCam magnitude \\
e\_m360mag & Error in m360mag \\
m444mag & F444W NIRCam magnitude \\
e\_m444mag & Error in m444mag
\enddata
\tablenotetext{a}{
(1) \citet{wei09};
(2) \citet{ing14};
(3) \citet{sle04};
(4) \citet{luh24o}.}
\tablecomments{
The table is available in its entirety in machine-readable form.}
\end{deluxetable}

\end{document}